\documentclass[prl,aps,twocolumn,showpacs]{revtex4}
\usepackage{bm}
\usepackage{epsfig}
\usepackage{amsmath}

\renewcommand{\v}[1]{{\bf #1}}
\newcommand{\bpm}{\begin{pmatrix}}
\newcommand{\epm}{\end{pmatrix}}
\newcommand{\ba}{\begin{eqnarray}}
\newcommand{\ea}{\end{eqnarray}}
\newcommand{\nn}{\nonumber \\}

\begin{document}

\title{Orbital Angular Momentum Origin of Rashba-type Surface Band Splitting}
%\title{Orbital Origin of the Chiral Angular Momentum Structure in Rashba Effect}
%\title{Dipolar Origin of Rashba-type Surface Band Splitting}

\author{Seung Ryong Park$^1$, Choong H. Kim$^2$, Jaejun Yu$^2$, Jung Hoon Han$^3$, Changyoung Kim$^4$}
\email[Electronic address:$~~$]{changyoung@yonsei.ac.kr}
\affiliation{$^1$Department of physics, University of Colorado at Boulder, Boulder, Colorado 80309, USA}
\affiliation{$^2$ Department of Physics and Astronomy, Seoul National University, Seoul 151-747, Korea}
\affiliation{$^3$Department of Physics and BK21 Physics Research Division, Sungkyunkwan University, Suwon 440-746, Korea}
\affiliation{$^4$Institute of Physics and Applied Physics, Yonsei University, Seoul 120-749, Korea}

\date{\today}

\begin{abstract}
We propose that existence of local orbital angular momentum (OAM) on the surfaces of high-$Z$ materials plays a crucial role in the formation of Rashba-type surface band splitting. Local OAM state in a Bloch wave function produces asymmetric charge distribution (electric dipole). Surface-normal electric field then aligns the electric dipole and results in chiral OAM states and the relevant Rashba-type splitting.  Therefore, the band splitting originates from electric dipole interaction, not from the relativistic Zeeman splitting as proposed in the original Rashba picture. The characteristic spin chiral structure of Rashba states is formed through the spin-orbit coupling and thus is a secondary effect to the chiral OAM. Results from first principles calculations on a single Bi layer under an external electric field verify the key predictions of the new model, including the direction of the spin texture which we predict to be in opposite direction to the conventional Rashba picture.
\end{abstract}
 \maketitle

%General introduction to Rashba effect
In the periodic band of a solid with inversion and time-reversal symmetries, Kramer's theorem mandates that each momentum state at $\v k$ be spin degenerate\cite{Kramers}. The spin degeneracy may be lifted, however, on surfaces of solids or interfaces of hetero-structures where the inversion symmetry is broken\cite{Nitta}. Lifting of the spin degeneracy due to inversion symmetry breaking (ISB) is usually called the Rashba effect\cite{Rashba}, which also entails a chiral spin structure along a constant energy contour as a consequence of the spin-momentum locking. Surface energy splitting and concomitant chiral spin structure have been experimentally observed on Au(111)\cite{LaShell,Reinert,Nicolay,WhoDidSpin}, Bi\cite{Bi}, Sb\cite{Sb}, and on some alloys as well\cite{alloy}. The effect has drawn additional attention recently due to its relevance for the surface states of topological insulators\cite{kane-reivew}.

%Problems in the Rashba effect
In spite of its phenomenological success, the Rashba Hamiltonian $H_\mathrm{R} = \lambda_\mathrm{R}\bm \sigma \cdot (\v k \times \hat{z})$, $\lambda_\mathrm{R}$=Rashba parameter, $\bm \sigma$=Pauli matrix, $\v k$=electron momentum and $\hat{z}$=surface normal, has several shortcomings when applied to solid surface phenomena. Estimates of Rashba energy $E_\mathrm{R}$ using the realistic work function at the surface gives $E_\mathrm{R}\sim 10^{-6}$ eV\cite{Petersen}, a value far too small to account for the observed energy splitting as large as a few hundred meV\cite{LaShell,Reinert,Nicolay,WhoDidSpin,Bi,Sb}. In addition, the Rashba picture has difficulty explaining the observation that the splitting increases with the atomic spin-orbit coupling (SOC) strength $\alpha$\cite{Reinert, Nicolay} because the strength of the surface electric field should not vary too much as a function of the atomic number\cite{workfunction}.

Several recent studies have presented improvements and/or alternatives to the original Rashba model. Within the tight-binding model, Petersen \textit{et al.} showed that the split energy indeed should be proportional, simultaneously, to $\alpha$ and surface potential gradient $\gamma$\cite{Petersen}. Even though not explicitly discussed in their work, the energy splitting also comes out to be proportional to the electron momentum $\v k$, which is essential in forming a Dirac cone-like dispersions. Some recent first-principles calculations show that the energy splitting is closely related to the electric potential energy of the asymmetric charge distribution at the surface\cite{Nagano,Yaji}. Orbital-mixing character of surface bands was emphasized by Bihlmayer \textit{et al}\cite{Bihlmayer}. Surface band splitting can be very large purely by band structure effect at specific wave vectors\cite{Emmanouil}. However, efforts to find the microscopic origin of the surface energy splitting and its consequences so far have been largely numerical, and do not seem to confront the underlying physical mechanism. The tight-binding model analysis in Ref. \onlinecite{Petersen} points out the simultaneous presence of SOC and surface electric field as key factors in the Rashba splitting, but does not incorporate the wave function asymmetry found in later band calculations. The physical origin of the chiral spin structure was not addressed in band calculation approaches\cite{ohgushi}, and one needs to resort to the original Rashba picture when the chiral spin structure is discussed. In addition, as we will discuss later, the direction of the spin texture in strong SOC cases predicted by the Rashba picture is opposite to what has been experimentally measured, which has been somehow unnoticed.

In this article, we argue that the most important aspect of the energy splitting and resulting chiral spin structure is the existence of orbital angular momentum (OAM) in high-$Z$ materials due to the strong atomic SOC. We further show that such an OAM state in combination with the electron momentum causes asymmetric charge distribution, which in turn determines not only the energy level but also the OAM direction through electric dipole interaction. The spin angular momentum (SAM) direction is determined by the OAM direction because of the strong SOC, forming a chiral SAM structure. Therefore, SAM texture is a secondary effect due to the OAM chiral structure, not the primary effect, contrary to what has been been viewed so far. In order to verify the new model, we also present first-principles calculation results on a single-layer Bi sheet under an external electric field. The results are consistent with the new model.

\begin{figure}
\centering \epsfxsize=7.5cm \epsfbox{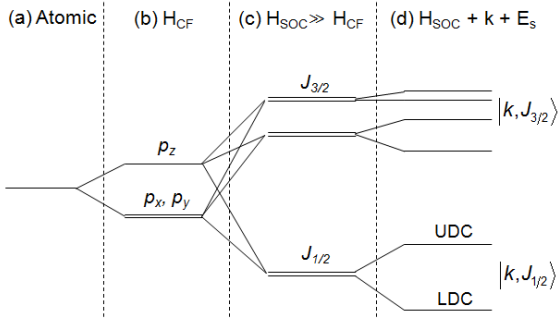} \caption{Energy levels of $p$ states (a) in an atom, (b) with a weak crystal field, (c) with a strong SOC and (d) in a tight bing state. H$_{CF}$ and E$_s$ are the crystal field Hamiltonian and surface electric field, respectively.}\label{fig1}
\end{figure}

We consider a system consisting of $p$-orbitals in this paper, which is of practical importance as the relevant surface bands of elements such as Bi, Sb, and Pb exhibiting Rashba-type splitting consist mainly of $p$-orbitals. Even the valence states of Au surface show $p$-orbital character due to strong mixing between $s$- and $p$-orbitals by surface electric field\cite{Nagano}.

%Figure 1 - reappearance of OAM
For an intuitive understanding, we consider a case in which the SOC is very large. Energy levels for such case are schematically shown in Fig. 1. Without SOC, OAM is quenched in the presence of crystal-field and, $p$-orbitals are the energy eigen-states as illustrated  in the figure. When SOC is turned on, the total angular momentum $\bf J$ eigen-states become the energy eigen-states instead. The most important aspect of this is that OAM comes back to the picture as a relevant low-energy degree of freedom as will be explained later. At the last stage of Fig. 1, the OAM in combination with the electron momentum $\v k$ causes asymmetric charge distribution or electric dipole moment whose direction is opposite for the states with opposite OAM directions. In the ISB circumstances such as on surfaces, the electric field splits the states through electric dipole interaction. Subsequently, upper and lower Dirac cone (UDC and LDC, respectively) states depicted in the figure are formed.

%Figure 2a
\begin{figure}
\centering \epsfxsize=7.5cm \epsfbox{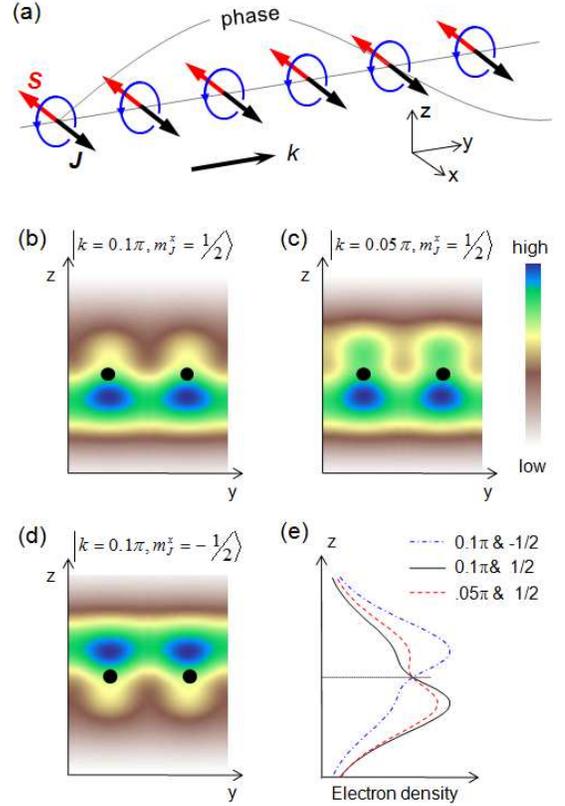} \caption{(a) A tight binding state built from $J_{1/2}$ states. Each $J_{1/2}$ state is the local state at each atom. Phase of each state is represented by the sinusoidal grey line. Electron density of a tight binding state integrated along x-direction for (b) $\v k=0.1\pi$ $\&$ $\hat{\bm n}=\hat{\bm x}$, (c) $\v k=0.05\pi$ $\&$ $\hat{\bm n}=\hat{\bm x}$, and (d) $\v k=0.1\pi$ $\&$ $\hat{\bm n}=-\hat{\bm x}$. The ratio between Bohr radius for the $p$-orbital and the lattice constant is taken to be $7/20$. Black dots in (b)-(d) indicate positions of atoms. (e) Integrated (along both the $x$- and $y$-directions) electron density as a function of $z$ for the three cases.}\label{fig2}
\end{figure}

The first step to understanding the process is to see how the asymmetric charge distribution is formed. Let's focus on the $J=1/2$ doublet

\ba |u\rangle &=& {1\over\sqrt{3}} \Bigl( |p_x
\!\downarrow \rangle + i |p_y \!\downarrow \rangle + |p_z \!\uparrow
\rangle \Bigr), \nn
|d \rangle &=& {1\over\sqrt{3}} \Bigl(|p_x \!\uparrow \rangle - i
|p_y \!\uparrow \rangle - |p_z \!\downarrow \rangle \Bigr),
\label{eq:J-half-basis}\ea
which can be used to form a coherent state $|\hat{\bm n} \rangle$ satisfying

\ba |\hat{\bm n} \rangle = \cos {\theta \over 2} |u \rangle +
e^{i\phi} \sin {\theta \over 2} |d \rangle, ~~ \v J \cdot \hat{\bm n}
|\hat{\bm n}\rangle = {1\over 2} |\hat{\bm n}\rangle . \ea
Here $\hat{\bm n}= (\sin \theta \cos \phi, \sin \theta \sin \phi,
\cos \theta)$ is the unit vector indicating the direction of the
total angular momentum average $\langle \v J \rangle$, $\v J = \v L
+ (1/2) \bm \sigma$. A Bloch eigenstate, characterized by both the
wave vector $\v k$ and the total angular momentum orientation $\hat{\bm n}$, can
be formed, $|\hat{\bm n}, \v k\rangle = N^{-1/2}\sum_{i} e^{i \v k
\cdot \v r_i}|\hat{\bm n}, \v r_i \rangle$, as schematically shown
in Fig. 2(a) for $\v J$ (thus $\v L$ too) parallel to the
$x$-direction. Here, $N$ is the number of sites and $|\hat{\bm n},
\v r_i \rangle$ refers to the Wannier state at the atomic site $\v
r_i$. One can show, as in Fig. 2, that the density distribution of
the Bloch state depends on both $\v k$ and $\hat{\bm n}$ now, and in
particular results in nonzero dipole moment that varies with them both.

%Figure 2(b). OAM can induce dipole moment
To demonstrate that OAM can induce nonzero dipole moments, we plot in Fig. 2(b)-(d) the density of the Bloch state for different $\v k$ and $\hat{\bm n}$ values. For simplicity, we use hydrogenic 2$p$ states as the orbital part of the Wannier state. The plotted densities are integrated electron densities of the Bloch state along the $x$-direction. It is clear from Fig. 2(b) that electron density is higher in the $-z$ region. This asymmetric electron distribution can be understood in the following way. One can view the tight-binding state as a superposition of a free electron and local OAM states. When the phase velocities of the electron and OAM states are pointing in the same direction ($-z$ region), there is a constructive interference which results in a higher electron density while opposite motions cause destructive interference, yielding lower electron densities ($+z$ region).

%Figure 2. Dipole moment proportional to L and k
Electron densities for other combinations of $\v k$ and $\hat{\bm n}$ values are plotted in Figs. 2(c) and 2(d). The charge density becomes less asymmetric as the momentum $\v k$ decreases from 0.1$\pi$ to 0.05$\pi$. It should eventually become symmetric at $\v k$=0 for an obvious reason. On the other hand, the density is higher in the $+z$ region when angular momentum is flipped to $\hat{\bm n}=-1$ state (Fig. 2(d)). The trend can be seen more clearly by comparing integrated electron densities for the three cases plotted in Fig. 2(e). One can intuitively understand that the resulting electric dipole moment $\v d$ should be proportional to $\v L \times \v k$.

%Figure 3. Energy split
The significance of the OAM induced electric dipole moment is that the surface electric field $\v E_s$, directed normal to the $xy$ plane, couples to the dipole moment and result in the splitting of energy among different $\v L$ (thus $\hat{\bm n}$ in equation (2)). To see this, we can compute dipole interaction energy $E_\mathrm{d}(\v k) = -eE_s \int d\v r~ z |\psi_{\hat{\bm n}, \v k} (\v r) |^2 $, $e$=electric charge, $E_s$=electric field strength, and $\psi_{\hat{\bm n}, \v k} (\v r)$=wave function of the Bloch state, to find
\ba E_\mathrm{d} (\hat{\bm n}, \v k) \propto -W \hat{z} \cdot
(\hat{\bm n} \times \v k)= W \v k \cdot (\hat{\bm n} \times \hat{z}
) \label{eq:Ed-on-n-k}\ea
in the small-$\v k$ limit, and $W=eEa$ denotes the work function of the surface. Our non-relativistic, electrostatic consideration leads to the correct form and the energy scale of the Rashba-type splitting.

\begin{figure}
\centering \epsfxsize=8.0cm \epsfbox{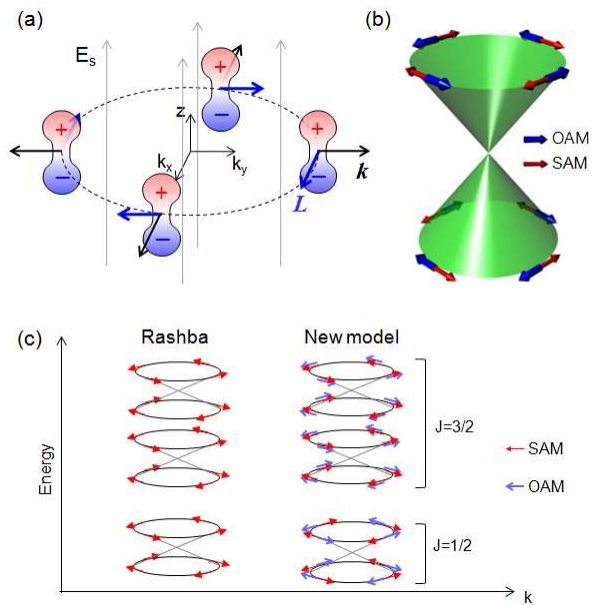} \caption{(a) Interaction of the electric dipole moment and surface electric field $E_s$, and resulting alignment of the OAM. The lobes schematically show the electron distribution of the tight binding state. This particular configuration represent the lower energy state of $J_{1/2}$. (b) Resulting SAM and OAM textures as well as the (Dirac) band dispersion. (c) Angular momentum structures predicted by the Rashba model (left) and new model (right).}\label{fig3}
\end{figure}

%Figure 3. Direction of L and chirality - figure 3(a) and 3(b)
An important consequence of the dipole interaction is that $\v L$ now has a preferred direction. The maximum/minimum energy is obtained for $\v L\times\hat{z}$ directed either parallel or anti-parallel to the $\v k$-vector (see Fig. \ref{fig3}(a)). Therefore, the direction of the $\v L$ is determined by the relationship $\v E_s \parallel \pm\v L\times\v k$, and it is the OAM that determines the locking of the angular momentum to the electron momentum. Due to the strong SOC, SAM $\v S$ is anti-aligned to $\v L$, which is a secondary effect. The resulting SAM and OAM structures for $J=1/2$ are illustrated in Fig. 3(b) which are compatible with earlier reports\cite{ParkCD}. Here, we point out that $\v E_s$ originates from the ISB, and therefore wave functions are determined by the ISB.

%Rashba vs new model - Figure 3(c)
There are two crucial differences between the SAM structures predicted by the two models. First of all, the SAM structure in the Rashba picture for strong SOC cases is found to be opposite to that in the new model shown in Fig. 3(b).  For an electron moving in the $x$-direction in Fig. 3(a), the effective magnetic field $\v B_{eff}=-(\hbar/mc^2)\v k\times \v E_s$ is along the $+y$-direction. For such magnetic field, spin along the $-y$-direction gives the lower energy. This results in clock-wise chiral spins for the lower Dirac cone, opposite to that of the new model as shown in Fig. 3(c). Experimental data from, for example, Au(111) surface states which correspond to $J=1/2$ states are consistent with the result of the new model. This problem in the Rashba picture has not been noticed so far. The other difference is that the SAM structures for the $J=1/2$ and $3/2$ cases are opposite in the new model. This is because OAM and SAM are directed parallel to each other for the $J=3/2$ states. On the other hand, the OAM structure is expected to be the same for all the Dirac cones (split bands near the $\Gamma$ point) as the Rashba Hamiltonian does not consider the OAM. The angular momentum structures predicted by the two models for an electric field along the $\hat z$ are schematically shown in Fig. 3(c). Confirmation of the SAM structures would be a definitive evidence for the new model.

\begin{figure}
\centering \epsfxsize=8.0cm \epsfbox{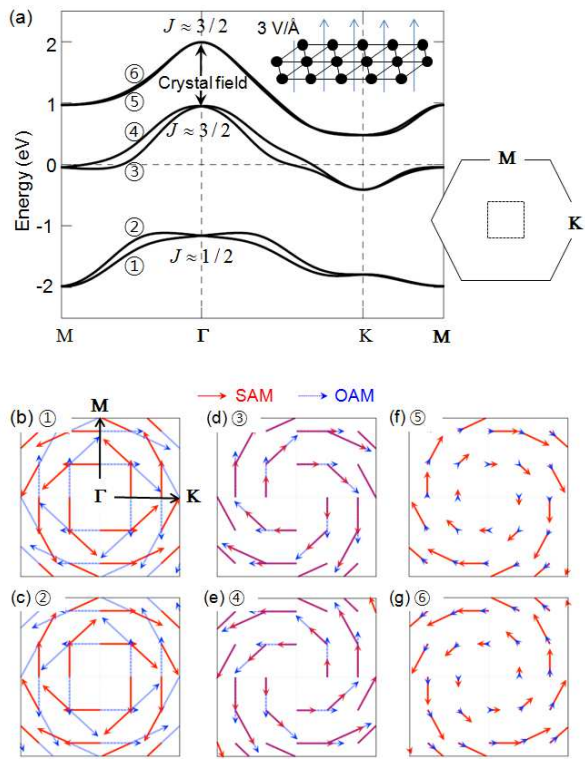} \caption{(a) Calculated band struture along the high symmetry directions for a single Bi layer in a triangular lattice under an external electric field of 3 $V/\mathring{A}$ along the $\hat z$ direction as shown in the inset. (b) - (g) SAM and OAM expectation values. Bands are numbered in panel (a). Red and blue arrows represent SAM and OAM. The $\v k$-space region for (b)-(g) is shown as the dotted square in the first Brillouin zone on the right-hand side of panel (a).}\label{fig4}
\end{figure}

%Test experiment - figure 4(a)
One may think about checking out the SAM structures on the known surface states such as Au(111) surface states. However, $J=3/2$ states are normally in the bulk states and cannot be separated from the bulk states. In addition, there could be a question on the direction of the surface electric field even though it is more likely out-of-surface direction. In order to circumvent these problems, we performed the first principles calculation on a single-layer of Bi in a triangular lattice under an external electric field. In this way, we simulate the surface with a definitive field direction and without bulk states (inset of Fig.
4(a)). The applied field (surface field in the $\hat z$ direction) is 3 $V/\mathring{A}$ which is a reasonable number considering the energy (work function) and length ($\sim$atomic size) scales.

For the density-functional theory (DFT) calculations within the local-density approximation, we used the DFT code, OpenMX,\cite{openmx} based on the linear-combination-of-pseudo-atomic-orbitals (LCPAO) method\cite{ozaki} and spin-orbit couplings were included via the norm-conserving, fully relativistic $j$-dependent pseudopotential scheme  in the non-collinear DFT formalism\cite{macdonald,bachelet,theurich}. LCPAO coefficients at specific $k$-points were used to calculate the SAM and OAM.

The resulting band structure is plotted in Fig. 4(a). The initially degenerate bands are split upon application of the field as expected. It should be noted that the top bands (5 and 6) primarily consist of $p_x$ and $p_y$ orbitals and are less susceptible to formation of OAM. This results in a much smaller splitting and thus supports our view that formation of OAM is essential in the energy splitting. On the other hand, split energies of other bands are similar to that of the Bi surface states even though a detailed level comparison is not meaningful.

%AM structures figures 4(c) - (g)
Calculated expectation values of SAM and OAM near the $\Gamma$ point are plotted in Figs. 4(b) - 4(g). We first note that OAM in bands 5 and 6 (where the splitting is very small) is much smaller than that in other bands, which again supports the new model. In comparing the results with the structures in Fig. 3(c), one finds that 1) the SAM direction for $J=1/2$ in the Rashba picture is indeed wrong, 2) SAM directions for $J=1/2$ and $3/2$ cases are opposite, suggesting that chiral structures are determined by the OAM, and 3) SAM and OAM are anti-parallel and parallel for $J=1/2$ and $3/2$ cases, respectively. These results are exactly what are expected in the new model, and thus conclusively prove our new understanding.

%Competition - atomic SOC parameter
What has not been discussed so far is the role of the atomic SOC parameter $\alpha$. Even though it is the OAM that determines the direction of the angular momentum structure, $\alpha$ still plays a crucial role in the Rashba-type splitting. Throughout the discussion given so far, we considered a large SOC, much larger than the crystal field. When $\alpha$ is very small, SAM and OAM do not have to be parallel or anti-parallel to each other. This means that OAM can point to the same direction for two different SAM directions and the state becomes spin degenerate (thus the splitting disappears).

%Dresselhaus effect
Finally, we argue that, even though ISB at surfaces and interfaces (Rashba effect) is considered, our model is quite general and has relevance to bulk cases (Dresselhaus effect)\cite{Dressel}. We may expect OAM in the bulk of high-$Z$ materials, provided the crystal field can be overcome by the strong SOC. In such cases, if there is ISB in the bulk like in zinc blend structures, electric dipole interaction can be significant. This may provide microscopic mechanism for the large bulk band spin splitting in III-V semiconductors\cite{De}. It was also reported that BiTeI bulk bands have a very large Rashba-type band splitting\cite{Ishizaka}. Since the Bi-layer is sandwiched between Te- and I-layers with negatively charged I-layer providing uniform electric field to the Bi layer, the situation is similar to the one in Fig. 4. Therefore, it will be interesting to check if OAM indeed exists in these materials.

\acknowledgements
This work was supported by the KICOS through Grant No. K20602000008 and by the Mid-career Researcher Program through NRF Grant No. 2010-0018092 (CK), No. 2011-0015631 (HJH) and R17-2008-033- 01000-0 (JY) funded by the MEST.

\end{document}